\documentclass[aps,pre,twocolumn,showcaps,superscriptaddress,showpacs,amsmath,amssymb]{revtex4}
\usepackage{graphicx}

\newcommand{\dd}{\mathrm{d}}

\newcommand{\mean}[1]{\langle #1 \rangle}

\newcommand{\IInt}[3]{\int_{#2}^{#3}\dd #1\;}

\newcommand{\ls}[1]{\langle #1|}
\newcommand{\rs}[1]{|#1\rangle}
\newcommand{\bracket}[2]{\langle #1|#2\rangle}

\renewcommand{\vec}[1]{\mathbf #1}

\newcommand{\al}{\alpha}

\newcommand{\lam}{\lambda}

\newcommand{\vhi}{\varphi}
\newcommand{\sig}{\sigma}
\newcommand{\im}{\mathrm{i}}
\newcommand{\ldf}{h}            % large deviation function
\newcommand{\eig}{\al}          % eigenvalue
\newcommand{\sm}{s_\mathrm{m}}
\newcommand{\dsm}{\dot s_\mathrm{m}}

\newcommand{\fc}{f_\mathrm{c}}

\begin{document}

\title{Large deviation function for entropy production in driven
  one-dimensional systems}

\author{Jakob Mehl}
\author{Thomas Speck}
\author{Udo Seifert}

\affiliation{{II.} Institut f\"ur Theoretische Physik,
Universit\"at Stuttgart, Pfaffenwaldring 57, 70550 Stuttgart,
Germany}

\begin{abstract}
  The large deviation function for entropy production is calculated for a
  particle driven along a periodic potential by solving a time-independent
  eigenvalue problem. In an intermediate force regime, the large deviation
  function shows pronounced deviations from a Gaussian behavior with a
  characteristic ``kink'' at zero entropy production. Such a feature can also
  be extracted from the analytical solution of the asymmetric random walk to
  which the driven particle can be mapped in a certain parameter range.
\end{abstract}


\maketitle

% ---- Introduction ----

The mathematical theory of large deviations is concerned with the exponential
decay of the probability of extreme events while the number of observations
grows~\cite{ellis}. In driven systems coupled to a heat reservoir, energy in
the form of heat is dissipated and therefore entropy in the surrounding medium
is produced. The large deviation function of the entropy production rate in
nonequilibrium steady states is a frequently studied quantity (see
Ref.~\cite{harr07} and references therein) for basically two reason. First,
since entropy is produced on average, the large deviation function captures
the asymptotically time-independent behavior of the probability distribution
for the entropy production. Second, the large deviation function exhibits a
special symmetry called fluctuation theorem or Gallavotti-Cohen
symmetry. First seen in computer simulations of a sheared
liquid~\cite{evan93}, the fluctuation theorem has been proven for both
deterministic thermostated dynamics~\cite{gall95,evan02} and stochastic
dynamics~\cite{kurc98,lebo99}.

Analytical solutions for the large deviation function exist only for a few
simple models~\cite{lebo99,visc06,wijl06}. Obtaining the large deviation
function over the full range from experimental data is a difficult task since
trajectories leading to negative entropy production are strongly suppressed
with increasing trajectory length (see, e.g., Ref.~\cite{cili04}). For a study
of the complete large deviation function one therefore has to rely on computer
simulations. To follow rare trajectories, different schemes have been proposed
and implemented~\cite{giar06,leco07,impa07}. All these approaches have in
common that they simulate trajectories from which the Legendre transform of
the large deviation function is determined. In contrast, in this Communication
we calculate numerically the Legendre transform directly as the lowest
eigenvalue of an evolution operator. We therefore reduce the problem of
determining a time-dependent probability distribution to solving a
time-independent eigenvalue problem.

For a simple paradigmatic system, we investigate a single driven colloidal
particle immersed in a fluid and trapped in a toroidal geometry by optical
tweezers such that it effectively moves in one dimension~\cite{fauc95,blic07}.
For short and intermediate times, the experimentally measured probability
distribution for the entropy production exhibits a detailed structure with
multiple peaks arising from the periodic nature of the
system~\cite{spec07}. As the observation time increases, the distribution
becomes more and more sharply peaked around its mean. Rare large deviations
from this mean are then governed by the large deviation function, which we
determine in this study. We then compare our results in a certain parameter
range to the analytically solvable model of the asymmetric random walk.

% ---- System ----

The colloidal particle is driven into a nonequilibrium steady state through a
constant force $f$. In addition, the particle moves within an external
periodic potential $V(\vhi)$, where $0\leqslant\vhi<2\pi$ is the angular
coordinate of the particle. The total force acting on the particle is
$F(\vhi)=-\partial_\vhi V(\vhi)+f$. The overdamped motion of the particle is
governed by the Langevin equation
\begin{equation}
  \label{eq:lang}
  \partial_t\vhi(t) = F(\vhi) + \zeta(t).
\end{equation}
The noise $\zeta$ represents the interactions of the particle with the fluid
and has zero mean and short-ranged correlations
$\mean{\zeta(t)\zeta(t')}=2\delta(t-t')$. Throughout the paper, we set
Boltzmann's constant to unity, leading to a dimensionless entropy. In
addition, we scale time and energy such that the bare diffusion coefficient
and the thermal energy become unity.

% ---- LDF ----

We are interested in the large deviation function of the entropy production
\begin{equation}
  \label{eq:ldf}
  \ldf(\sig) \equiv \lim_{t\rightarrow\infty}-\frac{1}{t}\ln p(\sm,t).
\end{equation}
The entropy $\sm$ produced in the heat bath during the time $t$ is a
stochastic quantity with probability distribution $p(\sm,t)$. The asymptotic
large fluctuations of $\sm$ are then given by
$p(\sm,t)\sim\exp[-\ldf(\sig)t]$, where $\sig\equiv(\sm/t)/\mean{\dsm}$ is the
dimensionless, normalized entropy production rate. We will not determine the
large deviation function $\ldf(\sig)$ directly through evaluating $p(\sm,t)$
but from the generating function 
\begin{equation}
  \label{eq:g}
  g(\vhi,\lam,t) \equiv \IInt{\sm}{-\infty}{+\infty} e^{-\lam\sm} 
  \rho(\vhi,\sm,t).
\end{equation}
Here, $\rho(\vhi,\sm,t)$ is the joint probability for the particle to be at an
angle $\vhi$ and to have produced an amount $\sm$ of entropy during the time
$t$. This generating function obeys the equation of motion $\partial_t g=\hat
L_\lam g$ with an operator $\hat L_\lam$ yet to be determined. We can then
expand $g$ into eigenfunctions $\psi_n(\vhi,\lam)$ determined from the
eigenvalue equation
\begin{equation}
  \label{eq:1}
  \hat L_\lam \psi_n(\vhi,\lam) = -\eig_n(\lam)\psi_n(\vhi,\lam).
\end{equation}
The lowest eigenvalue $\eig_0(\lam)$ determines the asymptotic time dependence
of the generating function $g\sim\exp[-\eig_0(\lam)t]$. In particular, the
mean entropy production rate is $\mean{\dsm}=\al_0'(0)$. The large deviation
function finally is the Legendre transform
\begin{equation}
  \label{eq:legendre}
  \ldf(\sig) = \eig_0(\lam^\ast) - \mean{\dsm}\sig\lam^\ast, \qquad
  \mean{\dsm}\sig = \eig'_0(\lam^\ast)
\end{equation}
of the cumulant generating function as can be shown by a saddle-point
integration. The large deviation function for the entropy production shows the
symmetry relation
\begin{equation}
  \label{eq:ft}
  \ldf(-\sig) = \ldf(\sig) + \mean{\dsm}\sig
\end{equation}
called fluctuation theorem~\cite{gall95,kurc98,lebo99}. If the fluctuation
theorem holds then the lowest eigenvalue exhibits an equivalent symmetry,
$\eig_0(\lam)=\eig_0(1-\lam)$. Hence, it is a symmetric function centered at
$\lam=1/2$~\cite{lebo99}.

In this approach, the asymptotic fluctuations of $\sm$ can be extracted from
the solution of the eigenvalue equation~\eqref{eq:1} for $n=0$. As an
advantage compared to following definition~\eqref{eq:ldf}, we do not have to
solve a time-dependent equation of motion for $p(\sm,t)$. Instead, the
information of the asymptotic fluctuations is contained in a time-independent
equation which we can tackle more easily.

% ---- specific dynamics ----

The entropy change along a single stochastic trajectory is defined as the
functional~\cite{seif05a}
\begin{equation*}
  \sm[x(\tau)] \equiv \IInt{\tau}{0}{t}F(x(\tau))\dot x(\tau) 
  = f(x-x_0) - \Delta V.
\end{equation*}
The time integration implies the introduction of a second angular coordinate
$x$ which takes into account the number of revolutions of the particle and
measures the total traveled distance in contrast to the bounded coordinate
$\vhi$. Since the terms involving $\Delta V$ and $x_0$ are bounded they will
not contribute to the entropy production rate in the limit of large
times. Hence, the expression for the entropy production in this limit
simplifies to $\sm\approx fx$.

The Fokker-Planck operator corresponding to the Langevin
equation~\eqref{eq:lang} reads
\begin{equation}
  \label{eq:L:0}
  \hat L_0 \equiv -\partial_\vhi(F-\partial_\vhi).
\end{equation}
In the next step, we want to obtain the evolution operator $\hat L$ for the
joint probability $\rho(\vhi,\sm,t)$ which obeys $\partial_t\rho=\hat L\rho$.
This operator is then converted to the sought-after evolution operator $\hat
L_\lam$ for the generating function~\eqref{eq:g}. The stochastic processes for
$\vhi$ and $x$ (and hence $\sm$) share the same noise. We can therefore
replace $\partial_\vhi\mapsto(\partial_\vhi+\partial_x)$ to obtain
\begin{equation}
  \label{eq:L}
  \hat L = \hat L_0 
  + (2\partial_\vhi - F)\partial_x + \partial_x^2.
\end{equation}
Differentiating Eq.~\eqref{eq:g} with respect to time and inserting the
operator~\eqref{eq:L} leads after partial integration to
\begin{equation}
  \label{eq:L:lam}
  \hat L_\lam = \hat L_0 + (2\partial_\vhi - F)f\lam + (f\lam)^2
\end{equation}
with vanishing boundary terms.

For the numerical evaluation, we represent the operator $\hat L_\lam$ as a
matrix through choosing a basis. To this end, we distinguish left sided
$\ls{k}$ from right sided $\rs{k}$ basis states. The basis must be complete
and orthonormal, $\bracket{k}{l}=\delta_{kl}$. Expanding the eigenfunctions
$\psi_n(\lam)=\sum_kc^{(n)}_k(\lam)\rs{k}$ into the basis, Eq.~\eqref{eq:1}
becomes
\begin{equation}
  \label{eq:6}
  \sum_{l=-\infty}^\infty L_{kl}c^{(n)}_l = -\eig_n(\lam)c^{(n)}_k, \qquad
  L_{kl} \equiv \bracket{k}{\hat L_\lam l}.
\end{equation}
Hence, we seek the lowest eigenvalue $\eig_0(\lam)$ of the matrix $\vec
L_\lam\equiv(L_{kl})$ where $\lam$ appears as a mere parameter. A suitable
choice for the basis is
\begin{equation}
  \label{eq:4}
  \bracket{k}{\vhi} = \frac{e^{-\im k\vhi}}{\sqrt{2\pi}}, \quad
  \bracket{\vhi}{k} = \frac{e^{+\im k\vhi}}{\sqrt{2\pi}}, \quad
  \IInt{\vhi}{0}{2\pi}\rs{\vhi}\ls{\vhi}=1
\end{equation}
due to the periodic nature of the system.

% ---- plot ----

\begin{figure*}[t]
  \centering
  \includegraphics{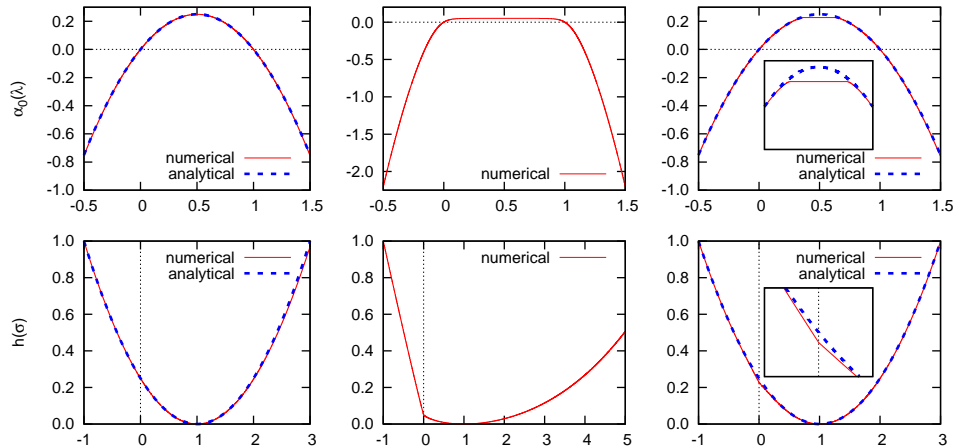}
  \caption{Eigenvalue $\eig_0(\lam)$ (top row) and large deviation function
    $\ldf(\sig)$ (bottom row) for the entropy production versus the force
    values $f=0.05$ (left), $f=4.05$ (center), and $f=100$ (right) for a
    potential depth $v_0=5$. The ordinates are in units of the mean production
    rate $\mean{\dsm}$. For small and large forces, the large deviation
    function and therefore the eigenvalues are almost parabolic. The
    corresponding analytical functions~\eqref{eq:quad} are shown for
    comparison. The insets in the right panels show the enlarged regions
    around $\lam=1/2$ and $\sig=0$, respectively.}
  \label{fig:heat}
\end{figure*}

% ---- potential ----

We now specialize our analysis to a cosine potential $V(\vhi)=v_0\cos\vhi$
introducing a second dimensionless parameter $v_0$. A straightforward
calculation shows that the matrix $\vec L_\lam$ becomes tridiagonal with
elements
\begin{gather*}
  L_{kk} = -(k-\im f\lam)^2 - \im f(k-\im f\lam), \\
  L_{k,k\pm1} = \pm \frac{v_0}{2}(k - \im f\lam).
\end{gather*}
We are not aware of an analytic solution for the eigenvalues of such a matrix.
However, by truncating the size of the matrix to some finite value, they can
easily be found numerically by standard algorithms. In Fig.~\ref{fig:heat}, we
show both $\eig_0(\lam)$ and the large deviation function $h(\sig)$ of the
entropy production. The fluctuation theorem~\eqref{eq:ft} is fulfilled as can
be seen immediately by the symmetry of $\eig_0(\lam)$. For large driving
forces $f\gg v_0$ as depicted in the right panels, both functions are almost
parabolic. In this case, the particle hardly ``feels'' the potential and the
mean velocity becomes $\mean{\dot x}\approx f$. Integrating over the angle
$\vhi$, the eigenvalue can then be read off from the operator~\eqref{eq:L:lam}
as
\begin{equation}
  \label{eq:quad}
  \eig_0(\lam) = \mean{\dsm}\lam(1-\lam), \quad
  \ldf(\sig) = (\mean{\dsm}/4)(\sig-1)^2
\end{equation}
with $\mean{\dsm}\approx f^2$. For small forces $f\ll v_0$ (left panels), the
particle remains mostly within one potential minimum and the mean rate becomes
exponentially small in the barrier height $2v_0$. In this case, the large
deviation function again approaches a parabola for which the
symmetry~\eqref{eq:ft} enforces the same functional form~\eqref{eq:quad} as in
the large force regime. The analytical functions~\eqref{eq:quad} are shown
together with the numerical curves for both small and large forces in
Fig.~\ref{fig:heat}.

\begin{figure}[b]
  \centering
  \includegraphics[width=\linewidth]{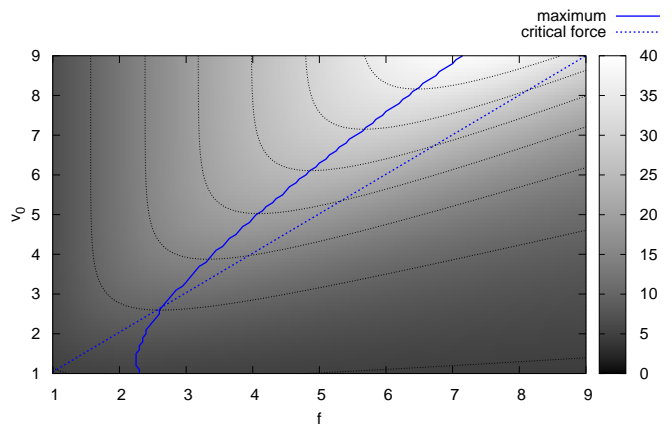}
  \caption{The ratio $\mean{\dsm}/\ldf(0)$ as a two dimensional contour plot
    for the parameters $f$ and $v_0$. The straight line is the critical force
    $f=v_0$. The thick line indicates the maximal value, i.e., it indicates
    the parameter pair for which the ``kink'' at $\sig=0$ in
    Fig.~\ref{fig:heat} becomes the most pronounced.}
  \label{fig:h0}
\end{figure}

Deviations from the simple Gaussian behavior show up in the right panels of
Fig.~\ref{fig:heat} even for surprisingly large forces. The eigenvalue
exhibits a flattening compared to the analytical curve around its center
$\lam=1/2$. This feature becomes more pronounced in the intermediate force
regime $f\simeq v_0$ (center panels). For the cosine potential $\fc=v_0$
corresponds to the critical force $\fc$, for which the barrier vanishes and
deterministic running solution for $\vhi(t)$ set in. In the large deviation
function $\ldf(\sig)$, this flattening corresponds to a ``kink'', an abrupt
albeit differentiable change around $\sig=0$. For an explanation of the
physical origin of this phenomenon note that all trajectories along which
$\sm$ grows slower than linearly in time are mapped onto $\sig=0$. If for a
large number of trajectories $\sm$ grows sublinearly, i.e., if $\sig=0$ has a
high probability density then $\ldf(0)$ becomes small. Due to the fluctuation
theorem~\eqref{eq:ft} and Eq.~\eqref{eq:legendre}, $\eig_0(1/2)=\ldf(0)$
always holds. Since the slope $\al_0'(0)=\al_0'(1)=\mean{\dsm}$ at $\lam=0$
and $\lam=1$ is fixed by the mean entropy production rate, for small
$\eig_0(1/2)$, i.e., for small $\ldf(0)$ the concave curve $\eig_0(\lam)$ must
become flat. In Fig.~\ref{fig:h0}, the ratio $\mean{\dsm}/\ldf(0)$ is plotted
together with the force curve $f^\mathrm{max}(v_0)$ for which
$\mean{\dsm}/\ldf(0)$ becomes maximal for fixed $v_0$. This curve indicating
the strongest ``kink'' is of the order of the critical force $\fc=v_0$. Hence,
it seems that in this force regime the particle disproportionately often stays
at or departs sublinearly from its initial position.

% ---- ARW ----

A similar ``kink'' in the large deviation function around $\sig=0$ can be
observed for the analytically solvable asymmetric random walk. The asymmetric
random walk is described by two rates $k^+$ and $k^-$ for a step forward and
backward, respectively. The entropy produced or annihilated in a single jump
is $b\equiv\ln(k^+/k^-)$~\cite{seif05,seif05a}. The random walker jumps $n^+$
steps forward and $n^-$ steps backward. The probability to have traveled
$n\equiv n^+-n^-$ steps in the forward direction during a time $t$ is known
analytically~\cite{vankampen},
\begin{equation}
  \label{eq:walk}
  p(n,t) = I_n(2\sqrt{k^+k^-}t)(k^+/k^-)^{n/2}e^{-(k^++k^-)t},
\end{equation}
where $I_n(z)$ is the modified Bessel function of the first kind of order $n$.
For the entropy production $\sm=bn$, the generating function~\eqref{eq:g}
becomes
\begin{equation*}
  \begin{split}
    g(\lam,t) &\equiv \sum_{n=-\infty}^\infty e^{-\lam bn}p(n,t) \\
    &= e^{-(k^++k^-)t} \sum_{n=-\infty}^\infty I_n(z) 
    \left[ \sqrt{k^+/k^-} e^{-\lam b} \right]^n.
  \end{split}
\end{equation*}
The sum can be evaluated using~\cite{abramowitz}
\begin{equation*}
  \sum_{n=-\infty}^\infty I_n(z)c^n = \exp\left[(z/2)(c+c^{-1})\right].
\end{equation*}
We thus obtain an exponentially decaying generating function
$g(\lam,t)=\exp[-\eig_0(\lam)t]$ with the single eigenvalue
\begin{equation}
  \label{eq:5}
  \eig_0(\lam) = k^+ \left[ 1 + e^{-b} - e^{-\lam b} - e^{-(1-\lam)b}
  \right]
\end{equation}
obeying the symmetry $\eig_0(\lam)=\eig_0(1-\lam)$ as expected~\footnote{The
  large deviation function for the entropy production of the asymmetric random
  walk has been obtained previously somewhat differently in
  Ref.~\cite{lebo99}.}. The curvature of the large deviation function
$h(\sig)$ at $\sig=0$ can now be obtained analytically as
\begin{equation}
  \label{eq:2}
  h''(0) = -\frac{[\al_0'(0)]^2}{\al_0''(1/2)}
  = \frac{k^+}{2}e^{-(3/2)b} \left( e^b-1 \right)^2.
\end{equation}
For fixed forward rate $k^+$, this expression diverges for $k^-\rightarrow 0$
($b\rightarrow\infty$), i.e., for vanishing backward steps.

% ---- mapping ----

\begin{figure}[t]
  \centering
  \includegraphics[width=\linewidth]{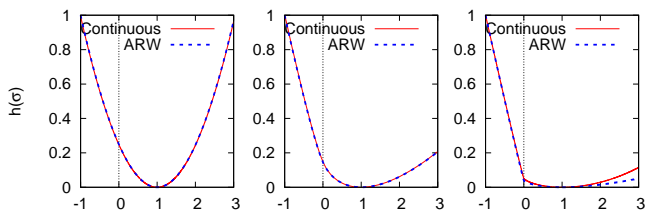}
  \caption{Comparison between the large deviation functions $\ldf(\sig)$ of
    the continuous dynamics and the asymmetric random walk (ARW). The
    parameters are: (left) $v_0=5$ and $f=0.05$, (center) $v_0=13$ and $f=1$,
    (right) $v_0=5$ and $f=4.05$.}
  \label{fig:cmp}
\end{figure}

In a parameter regime where the dynamics of the driven colloidal particle is
dominated by hopping events from one potential minimum to another, i.e., for
$f<v_0$, we can map the driven particle to the discrete asymmetric random
walk. The escape rate $k^+$ can be obtained by specializing the general
Kramers expression~\cite{hang90} as
\begin{equation}
  \label{eq:7}
  k^+ = \frac{v_0}{2\pi}\chi\exp\left\{ 
    -2v_0[\chi-(\pi/2)a+a\arcsin a] \right\}
\end{equation}
with $\chi\equiv\sqrt{1-a^2}$ depending on the two parameters $v_0$ and the
ratio $a\equiv f/v_0$. In Fig.~\ref{fig:cmp}, we compare the large deviation
functions of the continuous dynamics and the mapping to the corresponding
asymmetric random walk for three different parameter sets $v_0$ and $f$. For
small forces (left panel), the function $h(\sig)$ is a parabola as it should
be in such a linear response regime. Excellent agreement between the two
models is also obtained for deep potentials and larger forces (center panel)
for which $h(\sig)$ significantly deviates from a parabola but still shows no
``kink'' at $\sig=0$. Finally, in the right panel for $f$ approaching the
critical force, the mapping to the Kramers model breaks down as
expected. Still, both curves show a ``kink'' in this regime.

% ---- Conclusions ----

In summary, we have determined for a driven colloidal particle the large
deviation function of the entropy production by calculating the lowest
eigenvalue of the operator~\eqref{eq:L:lam}. We have used a numerical approach
to directly calculate the eigenvalue without simulating trajectories. This
approach can be extended to more complex systems with more than one degree of
freedom through choosing an appropriate basis. We have further compared our
results for a certain parameter range with a model where the large deviation
function and the eigenvalue can be obtained analytically. In both cases, the
large deviation function develops a ``kink'', an abrupt change around zero
entropy production. The question remains whether and to which extent this kink
will be a general feature also present in interacting systems.

We acknowledge financial support from the DFG.

% ---- References ----

\end{document}